# Note: "Lock in accelerometry" to follow sink dynamics in shaken granular matter

G. Sánchez-Colina[1], L. Alonso-Llanes[1], E. Martínez-Román[1], A. J. Batista-Leyva[2,1], C. Clement[3], C. Fliedner[3], R. Toussaint[3,a] and E. Altshuler[1,b].

[1] *"Henri Poincarè" Group of Complex Systems, Physics Faculty, University of Havana, 10400 Havana, Cuba.*
[2] *Instituto Superior de Tecnologías y Ciencias Aplicadas, La Habana, Cuba.*
[3] *Institut de Physique du Globe de Strasbourg (IPGS), Ecole et Observatoire des Sciences de la Terre (EOST), University of Strasbourg / CNRS, France*

[a] Electronic mail: renaud.toussaint@unistra.fr
[b] Electronic mail: ealtshuler@fisica.uh.cu

Understanding the penetration dynamics of intruders in granular beds is relevant not only for fundamental Physics, but also for geophysical processes and construction on sediments or granular soils in areas potentially affected by earthquakes. While the penetration of intruders in two dimensional (2D) laboratory granular beds can be followed using video recording, it is useless in three dimensional (3D) beds of non-transparent materials such as common sand. Here we propose a method to quantify the sink dynamics of an intruder into laterally shaken granular beds based on the temporal correlations between the signals from a *reference* accelerometer fixed to the shaken granular bed, and a *probe* accelerometer deployed inside the intruder. Due to its analogy with the working principle of a lock in amplifier, we call this technique Lock in accelerometry (LIA).

During Earthquakes, some soils can lose their ability to sustain shear and deform, causing subsidence and sometimes substantial building damage due to deformation or tumbling[1-3]. This soil liquefaction phenomenon happens typically in granular soils, wet or dry, and in saturated sedimentary soils. The stability under seismic waves was studied for sand[4] and dry granular soils[5], or for sediments[3].

It is usually performed by rheological tests in laboratory and numerical modeling[3]. The study of soil liquefaction using free surface boundary conditions, similar to the natural one, can benefit from direct measurement of the acceleration of soils submitted to oscillatory motion similar to those due to seismic waves.

The present note describes the development of a Lock-in accelerometry technique aimed at such measurements.

In the case of quasi-2D systems (like an intruder sinking into a Hele-Shaw cell filled with smaller sized grains) the penetration can be followed by using a video camera[6-8]. But video techniques are of no use to study the penetration into a 3D system of non-transparent grains, where the dynamics can be quite complex[9]. In those cases, wireless accelerometry constitutes a natural alternative that has been used in very few occasions, as far as we know[10-13].

However, it has never been used to quantify the penetration into horizontally shaken, sandy granular beds, where the vertical acceleration is small and then, difficult to follow. In this paper we propose a method to determine the sink time of an intruder into a horizontally shaken, fluidized granular bed based on the correlations between the signals of reference and probe accelerometers. The effectiveness of the system has been tested experimentally both in quasi-2D and in 3D systems.

Figure 1 shows our experimental setup. We use a Hele-Shaw cell that can oscillate laterally using an electromagnetic shaker with an amplitude of 1.5 cm and a maximum frequency of $\nu$ = 6 Hz. The cell consists in two vertical glasses separated by a gap of 21.4 ± 0.2 mm with wooden walls at the bottom and sides, filled up with multidisperse polymer particles, with average size of 0.7 ± 0.1 mm. At the bottom of the cell, a horizontal hose with 30 holes of 0.5 mm diameter each can inject air into the granular system at a flow rate ranging from 200 to 2200 cm$^3$/h. In the experiments presented here, the air flow range from 600 to 800 cm$^3$/h. Up to here, the system is analogous to previous ones reported in the literature[14]. However, in our case an intruder consisting in a squared parallelepiped 50 mm side, 19 mm thickness and a weight of 51 g is released on the free surface of the granular medium to study its sinking following a protocol to be explained later on. A digital camera *Hero 2* made by *GoPro* is fixed to the electromagnetic shaker, in such a way that it can take a video of the sinking process from an oscillating reference frame locked to the Hele-Shaw cell. Videos can be taken at a maximum rate of 120 frames per second (fps), with a resolution of 1920 × 1080 pixels.

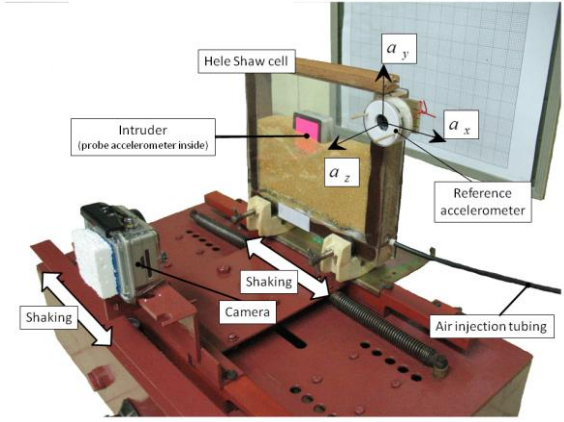

FIG. 1. Experimental setup for quasi-2D measurements. Both the Hele-Shaw cell and the camera are synchronously shaken in the lateral direction. Accelerometers attached to the Hele-Shaw cell and the intruder bring the key information to quantify the sink dynamics.

Finally, two 3-axis, wireless accelerometers are fixed to the Hele-Shaw cell (*Ref*), and inside the intruder (*Probe*), respectively. The *x*, *y* and *z* axis of the accelerometers are oriented as illustrated in Fig. 1 for the case of the reference accelerometer. Each 3-axis accelerometer has a resolution of 10$^{-4}$ g and is able to transmit data in real time at 2.4 GHz to a USB node on an external PC at a maximum data point rate of 120 Hz. This signal is used in the experiments here reported. The device[15] has a saturation acceleration of 8g (*g* = 9.81 m/s$^2$).

The protocol of a typical experiment can be described as follows. First, the granular bed is prepared by injecting air from the bottom, for 10 seconds. Then, the intruder is settled on the free surface of the granular bead. After activating the video camera and the accelerometers, the electromagnetic shaker and the air injection system are started at the same time, and turned off after 10 seconds, where the penetration process has ended. Then, the video and the acceleration records are analyzed.

As the horizontal acceleration is oriented in the *x* direction for both accelerometers, the reference (*R*) and the probe (*P*), we will compare both data sets $a_x^R, a_x^P$ using a modification of the Pearson correlation coefficient *r* aimed at decreasing the noise in the output. The Pearson coefficient for non centered data is defined as[16]:

$$r = \frac{\sum_{i=1}^{N} a_x^R(i) a_x^P(i)}{\left[ \sum_{i=1}^{N} \left(a_x^R(i)\right)^2 \sum_{i=1}^{N} \left(a_x^P(i)\right)^2 \right]^{\frac{1}{2}}} \quad (1)$$

where *i* represents the sampled time index and *N* is the total number of experimental samples during the experiment. We modified (1) by calculating the evolution of *r* within time intervals of size *D*, each one starting at moment *k* (so that *k* runs from 1 to *N* – *D*):

$$r(k) = \frac{\sum_{i=k}^{k+D} a_x^R(i) a_x^P(i)}{\left[ \sum_{i=k}^{k+D} \left(a_x^R(i)\right)^2 \sum_{i=k}^{k+D} \left(a_x^P(i)\right)^2 \right]^{\frac{1}{2}}} \quad (2)$$

The idea behind our experiment is that when the intruder is sinking, it cannot be tightly bounded to the granular mass, so there will be a delay between $a_x^R$ and $a_x^P$, giving a low correlation coefficient. On the contrary, when the intruder ends the sinking process, it starts to move synchronously with the reference frame, and so the correlation between $a_x^R$ and $a_x^P$ will be high.

Since lock in amplifiers work by comparing an oscillatory excitation (or reference) signal with an output signal from the probe, we call our technique lock in accelerometry.

However, in our case we can excite the system by a non-periodic vibration as could be expected in the case of earthquakes, for example.

An important property of our lock in method is that its results do not depend on changes in the relative orientation of the accelerometers caused by fixed misalignment between *Reference* and *Probe*. Also the correlation is not affected by slow rotations of the intruder. Both statements are demonstrated in Ref 17.

Figure 2 summarizes the main results from one sink experiment.

Figure 2(a) shows the time evolution of the penetration depth of the center of mass of the intruder, measured from a video recorded with the camera. Different stages are clearly identified. After an initial process of fast sinking, that takes around 1 s, the velocity of penetration decreases, and after around 4 s, a slow "creep" process takes place. After approximately 8 s, the sinking process ends as the intruder is confined by the "jammed" granular phase near the bottom of the cell.

In Figure 2(b) three accelerations are shown simultaneously: the horizontal acceleration, $a_x^R$, measured by the reference sensor (dark grey central line), the horizontal acceleration, $a_x^P$, measured with the probe sensor (black, upper line) and the vertical acceleration, $a_y^P$, measured by the probe sensor (light grey line, at the bottom). It is easy to see that the latter shows no evident features allowing us to follow the sinking process.

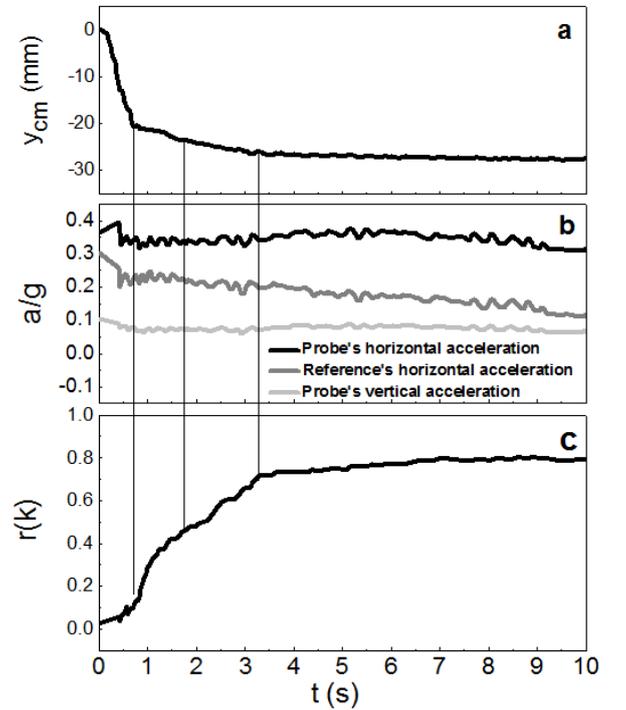

FIG 2: Time evolution of (a) penetration depth of the center of mass, (b) normalized accelerations and (c) correlation coefficient.

Finally, Figure 2(c) shows the time evolution of the correlation coefficient, $r(k)$, calculated by equation (2) for $D = 70$. It is easy to see that the value of $r$ reflects the main stages shown in Fig. 2(a): fast penetration between 0 and approximately 3 s, and "creep" motion between 3 s and approximately 8 s (a more detailed interpretation of the correlation curve will be the subject of future work).

In order to test our method in a 3D system, sink experiments were performed on a 3D printed cylinder of 44 mm diameter, 10 mm height (external dimensions) and 19.5 ± 0.5 g of mass – of effective density 1.28 g/cm$^3$. It was released in the upright position on the free surface of a granular bed contained into a box of 25 x 25 cm$^2$ base and 40 cm height, filled with polystyrene spheres of 80 μm diameter (monodisperse within a 1 %) and a bulk density of 1.05 g/cm$^3$ (Ugelstad spheres[18]). The box was horizontally shaken with an amplitude of 1.4 ± 0.1 cm and a frequency of 2.75 Hz. One accelerometer was fixed to the box, and the other to the sinking cylinder. The correlation between both was calculated during the sinking process, while the top of the cylinder (that never entered the granular bed) was observed, in the same spirit of the Hele Shaw experiment previously described. Here, the accelerometers where Analog Devices ADXL345 (Ref. 19) and the data was acquired by Arduino 2009 boards (Ref. 20).

The results of the experiment are shown in Fig. 3. Figure 3(a) shows the penetration depth as function of time in the 3D experiment as determined from the video. It is possible to see two stages in the process. Firstly a fast sinking, followed by a slower process until, finally, the depth stays constant.

Figure 3(b) shows the time evolution of the correlation coefficient between the accelerometers located into the intruder and the reference attached to the container in the same experiment. The beginning of time was taken before the start of the oscillations, and coincident with the origin in Fig. 3(a), when the correlation coefficient is 1 because both accelerometers are at rest. We selected $D = 40$ in eq. 2 (See Ref. [17] for a discussion about its influence on the results).

It is easy to see a steep decrement of $r$ as soon as the oscillations start: it indicates that the intruder moves freely in the first moments, almost uncorrelated with the sand. Then, the intruder sinks fast during approximately the first 2 seconds. As depth increases, it slows down probably when it reaches the "jammed" granular phase, and the correlation increases until approximately 0.8, where it stabilizes. This curve also shows clearly steps with different slopes, related with the two regions in Fig. 3(a).

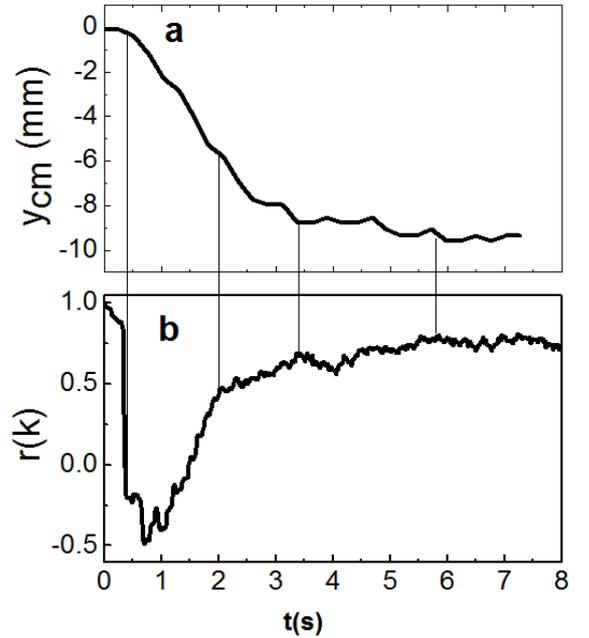

FIG. 3: Time evolution of the (a) penetration depth and (b) the correlation coefficient between the accelerometer of reference and the probe one in a 3D experiment.

In summary, we have demonstrated that our method is able to determine the time interval of "fast sinking" of an intruder into shaken granular beds for both quasi-2D and 3D systems. Further insight into the dynamics could be attained by studying the maximum normalized cross correlation between the accelerometers in time, and the phase lag between an intruder at different depths, and the granular surface.

In the present contribution we have studied dry materials, but the technique can also be used for wet granular matter. Finally, substituting the intruder by a solid rock and the granular bed by actual soil may expand the technique to measure, *in situ*, the rheological response of a soil during an earthquake.

We acknowledge support from Project 29942WL (Fonds de Solidarité Prioritaire France-Cuba), from the EU ITN FlowTrans, and from the Alsatian network REALISE.

**References**

[1] N.N. Ambraseys, Engineering seismology, Earthquake Eng. Structural Dynamics, 17, 1-105 (1988)

[2] National Research Council, Liquefaction of soils during Earthquakes, National Academy Press, Washington, DC (1985).

[3] C.Y. Wang and M. Manga, Earthquakes and water, Springer, Berlin Heidelberg (2014)

[4] J.B. Berril and R.O. Davis, Soils Found., 25, 106-118, (1985).

[5] E. Clement and J. Rajchenbach. Europhys. Lett. , 16 :133, (1991).

[6] I. Sánchez, G. Gutiérrez, I . Zuriguel and D. Maza, Phys. Rev. E **81**, 062301 (2010).

[7] M.J. Niebling, E.G. Flekkøy, K.J. Måløy, R. Toussaint, Phys. Rev. E 82, 051302, 2010.

[8] M.J. Niebling, E.G. Flekkøy, K.J. Måløy, R. Toussaint, Phys. Rev. E 82, 011301 (2010).

[9] T. Shinbrot and F. J. Muzio, Phys. Rev. Lett. **81**, 4365 (1998).

[10] F. Pacheco-Vázquez, G. A. Caballero-Robledo, J. M. Solano-Altamirano, E. Altshuler, A. J. Batista-Leyva and J. C. Ruiz-Suárez, Phys. Rev. Lett. **106**, 218001 (2011).

[11] H. Torres, A. González, G. Sánchez-Colina, J. C. Drake and E. Altshuler, Rev. Cub. Fis. **29**, 1E45 (2012).

[12] R. Zimmermann, L. Fiabane, Y. Gasteuil, R. Volk and J.-F. Pinton, New J. of Phys. **15,** 015018 (2013).

[13] E. Altshuler, H. Torres, A. González-Pita, G. Sánchez-Colina, C. Pérez-Penichet, S. Waitukaitis and R. Cruz, Geophys. Res. Lett.41, 3032 (2014)

[14] G. Metcalfe, S. G. K. Tennakoon, L. Kondic, D. G. Shaeffer and R. P. Behringer, Phys. Rev E. **65**, 031302 (2002)

[15] MMA7660FC ZSTAR3 accelerometer, details at www.freescale.com/zstar.

[16] F. Katagiri and J. Glazebrook, PNAS **100**, 10842 (2003).

[17] See supplementary material at [URL will be inserted by AIP Publishing].

[18] R. Toussaint, J. Akselvoll, E.G. Flekkøy, G. Helgesen and A.T. Skjeltorp, Phys. Rev. E 69, 011407 (2004).

[19] www.analog.com/static/imported-files/data-sheets/ADXL345.pdf

[20] www.arduino.cc/en/Main/arduinoBoardDuemilanove